\documentclass[aps,pra,reprint,showpacs,superscriptaddress,twocolumn]{revtex4-1}%
\usepackage{amssymb}
\usepackage{mathrsfs}
\usepackage{amsfonts}
\usepackage{graphicx}
\usepackage{amsmath}
\usepackage{dcolumn}
\usepackage{color}
\usepackage[colorlinks,linkcolor=blue,citecolor=blue,hyperindex,bookmarks=false,pdfstartview=FitH]{hyperref}
\begin{document}
\title{Nonreciprocal quantum interference and coherent photon routing in a three-port optomechanical system}
\author{Lei Du}
\affiliation{Beijing Computational Science Research Center, Beijing 100193, China}
\author{Yao-Tong Chen}
\affiliation{Center for Quantum Sciences, Northeast Normal University, Changchun 130117, China}
\author{Jin-Hui Wu}
\affiliation{Center for Quantum Sciences, Northeast Normal University, Changchun 130117, China}
\affiliation{State Key Laboratory of Quantum Optics and Quantum Optics Devices, Shanxi University, Taiyuan 030006, China}
\author{Yong Li}
\email{liyong@csrc.ac.cn}
\affiliation{Beijing Computational Science Research Center, Beijing 100193, China}
\affiliation{Synergetic Innovation Center for Quantum Effects and Applications, Hunan Normal University, Changsha 410081, China}

\begin{abstract}
We study the quantum interference between different weak signals in a three-port optomechanical system, which is achieved by coupling three cavity modes to the same mechanical mode. If one cavity serves as a control port and is perturbed continually by a control signal, nonreciprocal quantum interference can be observed when another signal is injected upon different target ports. In particular, we exhibit frequency-independent perfect blockade induced by the completely destructive interference over the full frequency domain. Moreover, coherent photon routing can be realized by perturbing all ports simultaneously, with which the synthetic signal only outputs from the desired port. We also reveal that the routing scheme can be extended to more-port optomechanical systems. The results in this paper may have potential applications for controlling light transport and quantum information processing.
\end{abstract}
\maketitle

\section{Introduction}\label{sec1}

Optomechanical systems provide a powerful platform for studying quantum mechanics because they may exhibit quantum effects on the macroscopic scale~\cite{rmp,OErev}. In optomechanics, mechanical motions can interact with optical modes in a large range of frequencies via radiation pressure. Such interactions, which are intrinsically nonlinear but can be linearized and enhanced via strong optical drivings, lead to a host of important applications such as ground-state cooling of mechanical resonators~\cite{cool1,cool2}, precise sensing~\cite{rmp,ms1,ms2,ms3,ms4}, and entanglement generation~\cite{entang1,entang2}. On one hand, one can manipulate mechanical motions and prepare quantum states for mechanical modes via optomechanical interactions~\cite{sqz1,sqz2,sqz3}. On the other hand, the mechanical motion can in turn affect the optical properties and thus results in some interesting quantum interference effects. For example, optomechanically induced transparency (OMIT)~\cite{OMIT1,OMIT2,OMIT3}, which is the optomechanical analogue of electromagnetically induced transparency (EIT)~\cite{EIT1,EIT2}, arises from the destructive interference between the probe signal and anti-Stokes field. Compared with simple two-mode cases, multimode optomechanical systems may exhibit richer quantum behaviors due to the multiple interference paths, such as topological energy transfer~\cite{topo}, coherent perfect absorption (CPA) and synthesis (CPS)~\cite{CPAagarwal,YanXB,DLepl}, ultra-narrow linewidth~\cite{DLoe}, and two-mode squeezing~\cite{tms1,tms2}.

Meanwhile, nonreciprocal optics has attracted vast interest due to its crucial applications in optical communication and quantum information processing. In particular, optomechanical systems provide a promising platform for realizing nonreciprocal devices. As is well known, the basic idea to realize nonreciprocity is to break the time-reversal symmetry of the system, which can be readily achieved in optomechanics without using the traditional magneto-optical effects~\cite{mag-op1,mag-op2}. In general, nonreciprocity can be realized by tuning the relative phase between different interference paths. With this method, optical isolators~\cite{hafezi,xxw1,xxw2,tl,dch1,yxbnew}, circulators~\cite{xxw1,circ,dch2}, and directional amplifiers~\cite{da1,da2,da3,da4} have been theoretically studied and experimentally realized. Inspired by the Sagnac effect, however, nonreciprocal optical transmissions can also be realized by spinning a whispering-gallery-mode (WGM) resonator~\cite{spin1,spin2}. With this technology, nonreciprocal photon blockade~\cite{jhblock} and phonon lasing~\cite{nrpl} have been achieved in succession. Most recently, Xu \emph{et al.} proposed a nonreciprocal transition scheme of nondegenerate energy levels based on reservoir engineering, which may serve as the kernel of a nonreciprocal single-photon transporter~\cite{xutrans}. In this context, it is natural to consider realizing more nonreciprocal optical phenomena via the seminal methods, such as nonreciprocal quantum interference.

In this paper, we propose a feasible scheme of nonreciprocal quantum interference and coherent photon routing based on a three-port optomechanical system, which consists of three indirectly coupled optical modes and one intermediate mechanical mode. By regarding one cavity as a control port and perturbing it continually by a control signal, the system may exhibit quite different quantum interferences if we inject another signal upon different target ports. The isolation ratio can be controlled by adjusting the phase difference between the control and target signals. With specific parameters, one can observe unidirectional or bidirectional transmission blockade over the full frequency domain. Moreover, coherent photon routing can be realized if we perturb all ports simultaneously. In this case, the synthetic signal only outputs from the desired port. Finally, we extend the results to a four-port optomechanical system to show the generalizability of our scheme.

\section{Models and equations}\label{sec2}

\begin{figure}[ptb]
\includegraphics[width=8.5cm]{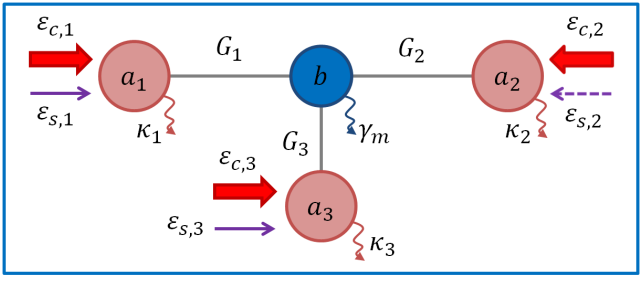} \caption{(Color online) Schematic illustration of the three-port optomechanical system under consideration. The cavity mode $a_{j}$ $(j=1,2,3)$ of decay rate $\kappa_{j}$ is driven by the red-sideband control field $\varepsilon_{c,j}$ and perturbed by the weak signal $\varepsilon_{s,j}$. Each cavity mode couples with the same mechanical mode $b$ of damping rate $\gamma_{m}$, with $G_{j}$ denoting the effective optomechanical coupling strength between $a_{j}$ and $b$.}
\label{fig1}%
\end{figure}

We consider in Fig.~\ref{fig1} a three-port optomechanical system, which consists of three cavity modes $a_{j}$ $(j=1,2,3)$ and one mechanical mode $b$. In this model, the resonance frequencies of the three cavity modes can be vastly different since there is no direct interaction between them, i.e., the mechanical mode serves as an intermediate bridge between the cavity modes. In addition, cavity mode $a_{j}$ of resonance frequency $\omega_{j}$ is driven by a strong control field of amplitude $\varepsilon_{c,j}$, phase $\vartheta_{j}$, and frequency $\omega_{c,j}$. In the interaction picture with respect to $H_{0}=\sum_{j}\omega_{c,j}a^{\dagger}_{j}a_{j}$, the Hamiltonian of the unperturbed system can be given by $(\hbar=1)$
\begin{eqnarray}
H_{u}&=&\omega_{m}b^{\dagger}b+\sum_{j=1}^{3}[\Delta_{j}a^{\dagger}_{j}a_{j}+g_{j}a^{\dagger}_{j}a_{j}(b^{\dagger}+b)\nonumber\\
&&+i\sqrt{\kappa_{\text{ex},j}}\varepsilon_{c,j}(e^{i\vartheta_{j}}a^{\dagger}-h.c.)],
\label{eq1}
\end{eqnarray}
where $\omega_{m}$ is the eigenfrequency of the mechanical mode and $\Delta_{j}=\omega_{j}-\omega_{c,j}$ is the detuning between cavity mode $a_{j}$ and control field $\varepsilon_{c,j}$. $g_{j}$ denotes the single-photon optomechanical coupling constant between $a_{j}$ and $b$. $\kappa_{\text{ex},j}$ is the external loss of cavity mode $a_{j}$. Note in our representation, the amplitudes of all control fields are real numbers.

Now we consider two scenarios under which the three-port optomechanical system is perturbed by weak signals in different ways. In both situations, we regard cavity $a_{3}$ as a control port and always inject upon it a control signal of amplitude $\varepsilon_{s,3}$. Cavities $a_{1}$ and $a_{2}$, however, serve as two target ports of our interest. For the first scenario, one can select cavity mode $a_{1}$ or $a_{2}$ to be perturbed by a target signal of amplitude $\varepsilon_{s,1}$ or $\varepsilon_{s,2}$ to investigate the interference phenomena in different directions. In this case, the Hamiltonian of signals (in the interaction picture) is given by
\begin{eqnarray}
H_{s,1(2)}&=&i\sqrt{\kappa_{\text{ex},1(2)}}\varepsilon_{s,1(2)}[e^{i\varphi_{1(2)}}e^{-i\Delta_{s,1(2)} t}a^{\dagger}_{1(2)}-h.c.]\nonumber\\
&&+i\sqrt{\kappa_{\text{ex},3}}\varepsilon_{s,3}(e^{i\varphi_{3}}e^{-i\Delta_{s,3} t}a^{\dagger}_{3}-h.c.)
\label{eq2}
\end{eqnarray}
with $\Delta_{s,j}=\omega_{s,j}-\omega_{c,j}$ being the detuning between signal $\varepsilon_{s,j}$ of frequency $\omega_{s,j}$ and control field $\varepsilon_{c,j}$. $\varphi_{j}$ is the phase of signal $\varepsilon_{s,j}$. In other words, one can investigate with $H_{s,1(2)}$ the quantum interference between the transmitted field from $a_{1(2)}$ to $a_{2(1)}$ and control signal $\varepsilon_{s,3}$. For the second scenario, we perturb all three cavity modes by weak signals of identical amplitude $(\varepsilon_{s,1}=\varepsilon_{s,2}=\varepsilon_{s,3}=\varepsilon_{s})$ and vanishing phase $(\varphi_{1}=\varphi_{2}=\varphi_{3}=0)$, i.e.,
\begin{equation}
H_{s,3}=\sum_{j=1}^{3}i\sqrt{\kappa_{\text{ex},j}}\varepsilon_{s}(e^{-i\Delta_{s,j} t}a^{\dagger}_{j}-h.c.).
\label{eq3}
\end{equation}
With Eq.~(\ref{eq3}), the quantum interference between the three signals can be investigated. In fact, we can provide a general Hamiltonian of signals
\begin{equation}
H_{s}=\sum_{j=1}^{3}i\sqrt{\kappa_{\text{ex},j}}\varepsilon_{s,j}(e^{i\varphi_{j}}e^{-i\Delta_{s,j} t}a^{\dagger}_{j}-h.c.)
\label{eq4}
\end{equation}
containing both situations mentioned above by choosing parameters properly. The system is thus described by the total Hamiltonian $H_{\text{tot}}=H_{u}+H_{s}$. Similar to the control fields, the amplitudes of weak signals in Eqs.~(\ref{eq2})-(\ref{eq4}) are also real.

Taking relevant dissipations and noises into account, one can attain the Heisenberg-Langevin equations of the mode operators according to Eqs.~(\ref{eq1}) and (\ref{eq4}), i.e.,
\begin{subequations}\label{eq5}
\begin{align}
\dot{a_{1}}=&-(i\Delta_{1}+\frac{\kappa_{1}}{2})a_{1}-ig_{1}(b^{\dagger}+b)a_{1}+\sqrt{\kappa_{\text{ex},1}}(\varepsilon_{c,1}e^{i\vartheta_{1}}\nonumber\\
&+\varepsilon_{s,1}e^{i\varphi_{1}}e^{-i\Delta_{s,1} t})+\sqrt{\kappa_{1}}a^{\text{in}}_{1},\\
\dot{a_{2}}=&-(i\Delta_{2}+\frac{\kappa_{2}}{2})a_{2}-ig_{2}(b^{\dagger}+b)a_{2}+\sqrt{\kappa_{\text{ex},2}}(\varepsilon_{c,2}e^{i\vartheta_{2}}\nonumber\\
&+\varepsilon_{s,2}e^{i\varphi_{2}}e^{-i\Delta_{s,2} t})+\sqrt{\kappa_{2}}a^{\text{in}}_{2},\\
\dot{a_{3}}=&-(i\Delta_{3}+\frac{\kappa_{3}}{2})a_{3}-ig_{3}(b^{\dagger}+b)a_{3}+\sqrt{\kappa_{\text{ex},3}}(\varepsilon_{c,3}e^{i\vartheta_{3}}\nonumber\\
&+\varepsilon_{s,3}e^{i\varphi_{3}}e^{-i\Delta_{s,3} t})+\sqrt{\kappa_{3}}a^{\text{in}}_{3},\\
\dot{b}=&-(i\omega_{m}+\frac{\gamma_{m}}{2})b-i(g_{1}a^{\dagger}_{1}a_{1}+g_{2}a^{\dagger}_{2}a_{2}+g_{3}a^{\dagger}_{3}a_{3})\nonumber\\
&+\sqrt{\gamma_{m}}b^{\text{in}},
\end{align}
\end{subequations}
where $\kappa_{j}=\kappa_{0,j}+\kappa_{\text{ex},j}$ is the total decay rate of $a_{j}$ with $\kappa_{0,j}$ being the intrinsic loss. $\gamma_{m}$ is the damping rate of $b$. $a_{j}^{\text{in}}$ $(b^{\text{in}})$ is the zero-mean-value vacuum (thermal) noise operator of $a_{j}$ $(b)$. For strong driving, each mode operator can be expressed as the sum of its mean value and the quantum fluctuation, i.e., $a_{j}=\alpha_{j}+\delta a_{j}$, $b=\beta+\delta b$. The mean values can be attained by setting all time derivatives in Eq.~(\ref{eq5}) vanishing and neglecting the weak signals and noises, i.e.,
\begin{subequations}\label{eq6}
\begin{align}
\alpha_{j}=\frac{2\sqrt{\kappa_{\text{ex},j}}\varepsilon_{c,j}e^{i\vartheta_{j}}}{\kappa_{j}+2i\Delta_{j}'},\label{eq6a}\\
\beta=\frac{-2i\sum_{j}g_{j}|\alpha_{j}|^{2}}{\gamma_{m}+2i\omega_{m}}\label{eq6b}
\end{align}
\end{subequations}
with $\Delta_{j}'=\Delta_{j}+g_{j}(\beta^{*}+\beta)$ being the effective detuning between cavity mode $a_{j}$ and control field $\varepsilon_{c,j}$ induced by the mechanical motion. With the mean values, the Heisenberg-Langevin equations of the quantum fluctuations can be linearized as
\begin{subequations}\label{eq7}
\begin{align}
\dot{\delta a_{1}}=&-(i\Delta_{1}'+\frac{\kappa_{1}}{2})\delta a_{1}-iG_{1}(\delta b^{\dagger}+\delta b)\nonumber\\
&+\sqrt{\kappa_{\text{ex},1}}\varepsilon_{s,1}e^{i\varphi_{1}}e^{-i\Delta_{s,1} t}+\sqrt{\kappa_{1}}a^{\text{in}}_{1},\\
\dot{\delta a_{2}}=&-(i\Delta_{2}'+\frac{\kappa_{2}}{2})\delta a_{2}-iG_{2}(\delta b^{\dagger}+\delta b)\nonumber\\
&+\sqrt{\kappa_{\text{ex},2}}\varepsilon_{s,2}e^{i\varphi_{2}}e^{-i\Delta_{s,2} t}+\sqrt{\kappa_{2}}a^{\text{in}}_{2},\\
\dot{\delta a_{3}}=&-(i\Delta_{3}'+\frac{\kappa_{3}}{2})\delta a_{3}-iG_{3}(\delta b^{\dagger}+\delta b)\nonumber\\
&+\sqrt{\kappa_{\text{ex},3}}\varepsilon_{s,3}e^{i\varphi_{3}}e^{-i\Delta_{s,3} t}+\sqrt{\kappa_{3}}a^{\text{in}}_{3},\\
\dot{\delta b}=&-(i\omega_{m}+\frac{\gamma_{m}}{2})\delta b-i\sum_{j=1}^{3}(G_{j}\delta a^{\dagger}_{j}\nonumber\\
&+G^{*}_{j}\delta a_{j})+\sqrt{\gamma_{m}}b^{\text{in}}
\end{align}
\end{subequations}
with $G_{j}=g_{j}\alpha_{j}$ being the effective optomechanical coupling strength. It is clear from Eq.~(\ref{eq6a}) that the modulus and phase of $G_{j}$ can be controlled by adjusting the amplitude $\varepsilon_{c,j}$ and phase $\vartheta_{j}$ of the control field. In the following, we assume $G_{j}=|G_{j}|e^{i\theta_{j}}$ with $\theta_{j}$ being its phase.

In this paper, we assume that all three cavity modes are driven on the mechanical red sidebands with $\Delta_{1}'=\Delta_{2}'=\Delta_{3}'=\omega_{m}$, and our system works in the resolved sideband regime with $\omega_{m}\gg\{\kappa_{j},\,\gamma_{m},\,G_{j}\}$. With these assumptions, one can perform the transformation
\begin{subequations}\label{eq8}
\begin{align}
&\delta a_{j}\rightarrow\delta a_{j}e^{-i\Delta_{j}'t},\,\,\delta a_{j}^{\text{in}}\rightarrow\delta a_{j}^{\text{in}}e^{-i\Delta_{j}'t},\\
&\delta b\rightarrow\delta be^{-i\omega_{m}t},\,\,\delta b^{\text{in}}\rightarrow\delta b^{\text{in}}e^{-i\omega_{m}t}
\end{align}
\end{subequations}
and thus simplify Eq.~(\ref{eq7}) into
\begin{subequations}\label{eq9}
\begin{align}
\dot{\delta a_{1}}=&-\frac{\kappa_{1}}{2}\delta a_{1}-iG_{1}\delta b+\sqrt{\kappa_{\text{ex},1}}\varepsilon_{s,1}e^{i\varphi_{1}}e^{-i\xi_{s,1} t}\nonumber\\
&+\sqrt{\kappa_{1}}a^{\text{in}}_{1},\\
\dot{\delta a_{2}}=&-\frac{\kappa_{2}}{2}\delta a_{2}-iG_{2}\delta b+\sqrt{\kappa_{\text{ex},2}}\varepsilon_{s,2}e^{i\varphi_{2}}e^{-i\xi_{s,2} t}\nonumber\\
&+\sqrt{\kappa_{2}}a^{\text{in}}_{2},\\
\dot{\delta a_{3}}=&-\frac{\kappa_{3}}{2}\delta a_{3}-iG_{3}\delta b+\sqrt{\kappa_{\text{ex},3}}\varepsilon_{s,3}e^{i\varphi_{3}}e^{-i\xi_{s,3} t}\nonumber\\
&+\sqrt{\kappa_{3}}a^{\text{in}}_{3},\\
\dot{\delta b}=&-\frac{\gamma_{m}}{2}\delta b-i\sum_{j=1}^{3} G^{*}_{j}\delta a_{j}+\sqrt{\gamma_{m}}b^{\text{in}},
\end{align}
\end{subequations}
where $\xi_{s,j}=\Delta_{s,j}-\Delta_{j}'\approx\omega_{s,j}-\omega_{j}$ is the detuning between signal $\varepsilon_{s,j}$ and cavity mode $a_{j}$. In the following of this paper, we only focus on the case of $\xi_{s,1}=\xi_{s,2}=\xi_{s,3}\equiv\xi$.

To study the steady-state optical response to the weak signals, we can express the steady-state solutions to Eq.~(\ref{eq9}) as $\langle \delta o\rangle=o_{-}e^{-i\xi t}+o_{+}e^{i\xi t}$ $(o=a_{j},\,b)$. Note the anti-Stokes component $o_{-}$ corresponds to the signal frequency $\omega_{s,j}$ while the Stokes component $o_{+}$ corresponds to the frequency $2\omega_{c,j}-\omega_{s,j}$ arising from a nonlinear wave mixing process.

For the first scenario, when the target signal is injected upon cavity mode $a_{1}$ $(\varepsilon_{s,2}=0)$, the anti-Stokes component of cavity mode $a_{2}$ is attained as
\begin{equation}
a_{2-}=\frac{-\sqrt{\kappa_{\text{ex},1}}f_{3}G_{1}^{*}G_{2}\varepsilon_{s,1}-\sqrt{\kappa_{\text{ex},3}}f_{1}G_{2}G_{3}^{*}\varepsilon_{s,3}e^{i\phi}}{D},
\label{eq10}
\end{equation}
where $D=f_{1}f_{2}f_{3}h+f_{2}f_{3}|G_{1}|^{2}+f_{1}f_{3}|G_{2}|^{2}+f_{1}f_{2}|G_{3}|^{2}$, $f_{j}=\kappa_{j}/2-i\xi$, and $h=\gamma_{m}/2-i\xi$. $\phi=\varphi_{3}-\varphi_{1}$ is the phase difference between signals $\varepsilon_{s,1}$ and $\varepsilon_{s,3}$, which is determined by redefining all quantum fluctuations as $\delta o\rightarrow\delta oe^{i\varphi_{1}}$. According to the input-output relation~\cite{walls}, the anti-Stokes component of the output field of cavity mode $a_{2}$ satisfies
\begin{equation}
a_{2,\text{out}-}=\sqrt{\kappa_{\text{ex},2}}a_{2-}.
\label{eq11}
\end{equation}
In this case, the transmission coefficient of the target signal can be defined as $t_{1\rightarrow2}=a_{2,\text{out}-}/a_{1,\text{in}-}$ with $a_{1,\text{in}-}=\varepsilon_{s,1}$. On the other hand, when the target signal is injected upon cavity mode $a_{2}$ $(\varepsilon_{s,1}=0)$, we have
\begin{equation}
a_{1-}=\frac{-\sqrt{\kappa_{\text{ex},2}}f_{3}G_{2}^{*}G_{1}\varepsilon_{s,2}-\sqrt{\kappa_{\text{ex},3}}f_{2}G_{1}G_{3}^{*}\varepsilon_{s,3}e^{i\phi}}{D}
\label{eq12}
\end{equation}
with $\phi=\varphi_{3}-\varphi_{2}$ in this case and other symbols being the same as those in Eq.~(\ref{eq10}). Similarly, the transmission coefficient of the target signal can be defined as $t_{2\rightarrow1}=a_{1,\text{out}-}/a_{2,\text{in}-}$ with $a_{1,\text{out}-}=\sqrt{\kappa_{\text{ex},1}}a_{1-}$ and $a_{2,\text{in}-}=\varepsilon_{s,2}$.

For the second scenario, with the assumption $\varepsilon_{s,1}=\varepsilon_{s,2}=\varepsilon_{s,3}=\varepsilon_{s}$ and $\varphi_{1}=\varphi_{2}=\varphi_{3}=0$, we attain
\begin{subequations}\label{eq13}
\begin{align}
&\frac{a_{1-}}{\varepsilon_{s}}=\frac{\sqrt{\kappa_{\text{ex},1}}M_{1}-\sqrt{\kappa_{\text{ex},2}}f_{3}G_{1}G_{2}^{*}-\sqrt{\kappa_{\text{ex},3}}f_{2}G_{1}G_{3}^{*}}{D},\\
&\frac{a_{2-}}{\varepsilon_{s}}=\frac{\sqrt{\kappa_{\text{ex},2}}M_{2}-\sqrt{\kappa_{\text{ex},1}}f_{3}G_{2}G_{1}^{*}-\sqrt{\kappa_{\text{ex},3}}f_{1}G_{2}G_{3}^{*}}{D},\\
&\frac{a_{3-}}{\varepsilon_{s}}=\frac{\sqrt{\kappa_{\text{ex},3}}M_{3}-\sqrt{\kappa_{\text{ex},1}}f_{2}G_{3}G_{1}^{*}-\sqrt{\kappa_{\text{ex},2}}f_{1}G_{3}G_{2}^{*}}{D}
\end{align}
\end{subequations}
with $M_{1}=f_{2}f_{3}h+f_{3}|G_{2}|^{2}+f_{2}|G_{3}|^{2}$, $M_{2}=f_{1}f_{3}h+f_{3}|G_{1}|^{2}+f_{1}|G_{3}|^{2}$, and $M_{3}=f_{1}f_{2}h+f_{2}|G_{1}|^{2}+f_{1}|G_{2}|^{2}$. In this way, we can define the normalized output energy~\cite{YanXB} of cavity mode $a_{j}$ as $S_{j}=|a_{j,\text{out}-}/a_{j,\text{in}-}|^{2}=|\sqrt{\kappa_{\text{ex},j}}a_{j-}/\varepsilon_{s}-1|^{2}$.

\section{Nonreciprocal quantum interference}\label{sec3}

\begin{figure}[ptb]
\includegraphics[width=8.5cm]{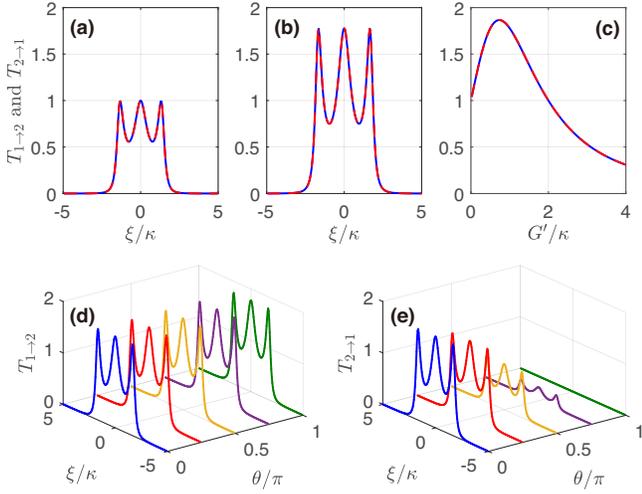} \caption{(Color online) Transmission rates $T_{1\rightarrow2}$ (blue solid) and $T_{2\rightarrow1}$ (red dashed) versus $\xi$ with $G'=0$ (a), versus $\xi$ with $G'=\kappa$ (b), and versus $G'$ with $\xi=0$ (c). Transmission rates $T_{1\rightarrow2}$ (d) and $T_{2\rightarrow1}$ (e) versus $\xi$ with different values of $\theta$. We assume $\theta=0$ in (a)-(c) while $G'=\kappa$ in (d) and (e). Other parameters are $\gamma_{m}=10^{-3}\kappa$, $G=\kappa$, $\eta=1$, and $\phi=0$.}
\label{fig2}%
\end{figure}

In this section, we consider the first scenario with $\kappa_{1}=\kappa_{2}=\kappa_{3}=\kappa$ $(f_{1}=f_{2}=f_{3}=f=\kappa-i\xi)$ for simplicity. As mentioned in Sec.~\ref{sec2}, the modulus and phase of $G_{j}$ can be readily controlled by adjusting $\varepsilon_{c,j}$ and $\vartheta_{j}$, respectively. Thus we assume the moduli as $|G_{1}|=|G_{2}|=G$, $|G_{3}|=G'$ and the phases as $\theta_{1}=\theta_{3}=0$, $\theta_{2}=\theta$. Moreover, we consider in this paper the so-called overcoupled regime~\cite{over1,over2}, i.e., $\kappa_{\text{ex},j}=\kappa_{j}=\kappa$. In this way, the transmission coefficients can be simplified as
\begin{equation}
t_{1\rightarrow2}=\frac{-\kappa(G^{2}+GG'\eta e^{i\phi})e^{i\theta}}{D'},
\label{eq14}
\end{equation}
and
\begin{equation}
t_{2\rightarrow1}=\frac{-\kappa(G^{2}e^{-i\theta}+GG'\eta e^{i\phi})}{D'}
\label{eq15}
\end{equation}
with $\eta=\varepsilon_{s,3}/\varepsilon_{s,1(2)}$ and $D'=f(fh+2G^{2}+G'^{2})$.

We first show in Figs.~\ref{fig2}(a)-\ref{fig2}(c) the impact of the third (control) port on the transmission rates $T_{j\rightarrow j'}=|t_{j'\rightarrow j}|^{2}$. In the absence of cavity $a_{3}$ $(G'=0)$, typical OMIT with three transparency windows can be observed, as shown in Fig.~\ref{fig2}(a). Such a phenomenon is reciprocal in both directions because the Hamiltonian is time-reversal symmetric in this case. Figure~\ref{fig2}(b) shows that in the presence of cavity $a_{3}$ and with proper parameters, the transmitted field is amplified due to the constructive interference between different transmission paths, i.e., $a_{1(2)}\rightarrow b \rightarrow a_{2(1)}$ and $a_{3}\rightarrow b \rightarrow a_{2(1)}$. The interference effect can be tuned by adjusting the effective optomechanical coupling strength between cavity $a_{3}$ and the mechanical mode $b$. As shown in Fig.~\ref{fig2}(c), one can observe constructive interference with $0<G'<2\kappa$ but destructive interference with $G'>2\kappa$. Note completely destructive interference requires a quite large $G'$ with which the system may enter the bistable region (not shown here).

It is clear from Eqs.~(\ref{eq14}) and (\ref{eq15}) that nonreciprocal quantum interference can be achieved by adjusting the phase difference $\theta$ ($T_{2\rightarrow 1}$ is sensitive to $\theta$ while $T_{1\rightarrow 2}$ is $\theta$-independent). This can be verified by the numerical results shown in Figs.~\ref{fig2}(d) and \ref{fig2}(e). We find that by changing $\theta$ from $0$ to $\pi$, $T_{1\rightarrow 2}$ remains invariable while $T_{2\rightarrow 1}$ decreases gradually. In particular, $T_{2\rightarrow 1}$ vanishes over the full frequency domain with $\theta=\pi$. We refer to such a phenomenon as \emph{frequency-independent perfect blockade} (FIPB). In this case, one can achieve within a large frequency range constructive interference in one direction but destructive interference in the opposite direction. Similarly, a contrary nonreciprocal phenomenon ($T_{2\rightarrow 1}\neq0$ and $T_{1\rightarrow 2}\equiv0$) can be achieved by assuming $\theta_{2}=\theta_{3}=0$ and $\theta_{1}=\theta$ instead with $\theta\neq2n\pi$ $(n=0,\,\pm1,\,...)$.

\begin{figure}[ptb]
\includegraphics[width=8.5cm]{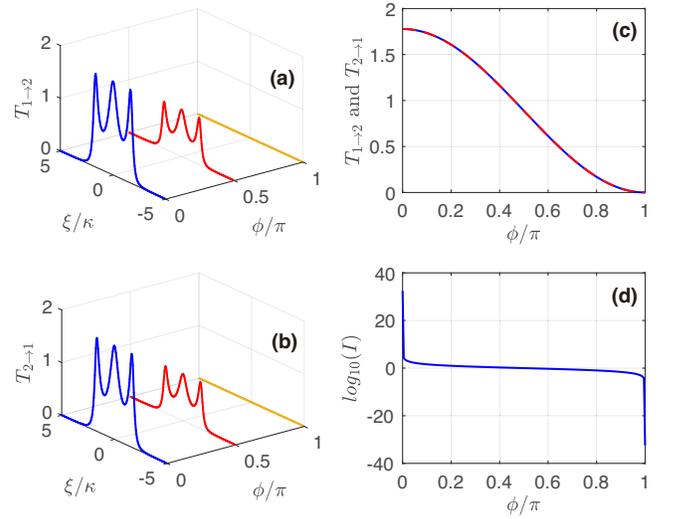} \caption{(Color online) Transmission rates $T_{1\rightarrow2}$ (a) and $T_{2\rightarrow1}$ (b) versus $\xi$ with different values of $\phi$. (c) Transmission rates $T_{1\rightarrow2}$ (blue solid) and $T_{2\rightarrow1}$ (red dashed) versus $\phi$ with $\xi=0$. (d) Isolation ratio versus $\phi$ with $\xi=0$. We assume $\theta=0$ in (a)-(c) and $\theta=\pi$ in (d). Other parameters are $\gamma_{m}=10^{-3}\kappa$, $G=G'=\kappa$, and $\eta=1$.}
\label{fig3}%
\end{figure}

The nonreciprocal phenomenon, however, is not available by only adjusting the phase difference $\phi$ in the case of $\theta=2n\pi$. As shown in Figs.~\ref{fig3}(a) and \ref{fig3}(b), $T_{1\rightarrow2}$ and $T_{2\rightarrow1}$ are always the same with the peak values decreasing gradually as $\phi$ increases from $0$ to $\pi$. In the case of $\phi=\pi$, both $T_{1\rightarrow2}$ and $T_{2\rightarrow1}$ vanish over the full frequency domain, implying that bidirectional FIPB is achieved. This result can also be understood from Eqs.~(\ref{eq14}) and (\ref{eq15}) that $T_{1\rightarrow2}\equiv T_{2\rightarrow1}$ in the case of $\theta=2n\pi$ while the phase difference $\phi$ only contributes to the magnitudes of the transmission rates. In Fig.~\ref{fig3}(c), we find that the transmission rates can be tuned continuously by adjusting $\phi$. Compared with the control scheme in Fig.~\ref{fig2}(c), the scheme here shows the advantage in terms of keeping the system away from its bistable region. Moreover, we point out that the isolation ratio $I=T_{1\rightarrow2}/T_{2\rightarrow1}$ can be tuned flexibly via $\phi$ in the nonreciprocal case, as shown in Fig.~\ref{fig3}(d). For $\theta=\phi=\pi$, we can also observe the nonreciprocal phenomenon ($T_{2\rightarrow 1}\neq0$ and $T_{1\rightarrow 2}\equiv0$) contrary to that shown in Figs.~\ref{fig2}(d) and \ref{fig2}(e).

\begin{figure}[ptb]
\includegraphics[width=8.5cm]{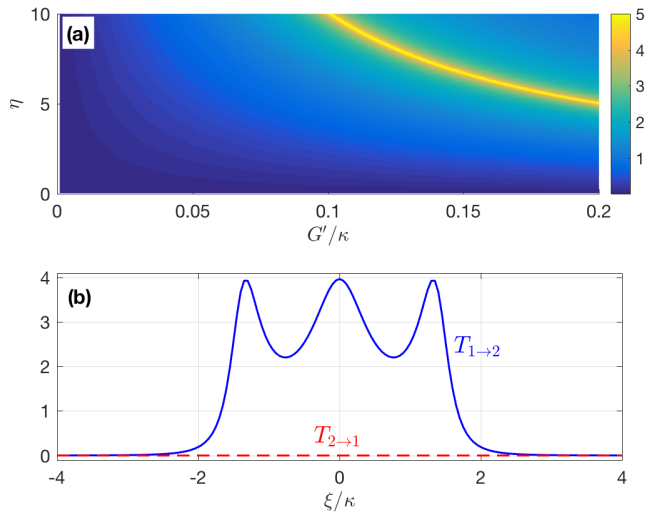} \caption{(Color online) (a) Logarithm of the isolation ratio $log_{10}(I)$ versus $G'$ and $\eta$ with $\xi=0$. (b) Transmission rates $T_{1\rightarrow2}$ (blue solid) and $T_{2\rightarrow1}$ (red dashed) versus $\xi$ with $G'=0.1\kappa$ and $\eta=10$. Other parameters are $\gamma_{m}=10^{-3}\kappa$, $G=\kappa$, $\theta=\pi$, and $\phi=0$.}
\label{fig4}%
\end{figure}

Note we have assumed $\varepsilon_{s,1(2)}=\varepsilon_{s,3}=\varepsilon_{s}$ $(\eta=1)$ in Figs.~\ref{fig2} and \ref{fig3} for simplicity. It is worth pointing out, however, that the nonreciprocity can be enhanced accompanied with FIPB persisting in one direction by adjusting $\varepsilon_{s,3}$ and $G'$ properly. According to Eqs.~(\ref{eq14}) and (\ref{eq15}), $G'\eta=G$ is the necessary condition of FIPB $(T_{2\rightarrow1}=0)$ in the case of $\theta=\pi$ and $\phi=0$. With this condition, one can enhance the transmission in the opposite direction $(T_{1\rightarrow2})$ by decreasing $G'$. Figure~\ref{fig4}(a) shows the logarithm of the isolation ratio $log_{10}(I)$ as a function of $G'$ and $\eta$, from which our analytical analysis is verified. As an example, we plot in Fig.~\ref{fig4}(b) the transmission rates versus the detuning $\xi$ with $G'=0.1\kappa$ and $\eta=10$. In this case, the transmission rate in one direction is increased by more than double while that in the opposite direction remains vanishing over the full frequency domain. Furthermore, we point out that the nonreciprocity can be significantly enhanced by driving cavity $a_{3}$ on the blue mechanical sideband (not shown here).

\section{Coherent photon routing}\label{sec4}

Now we consider the second scenario with $\varepsilon_{s,1}=\varepsilon_{s,2}=\varepsilon_{s,3}=\varepsilon_{s}$ and $\varphi_{1}=\varphi_{2}=\varphi_{3}=0$. For simplicity, we still assume $\kappa_{\text{ex},j}=\kappa_{j}=\kappa$ in this case. Moreover, we assume $|G_{3}|=G$ and $\theta_{3}=0$ ($G_{3}$ is purely real) without loss of generality.

\begin{figure}[ptb]
\includegraphics[width=8.5cm]{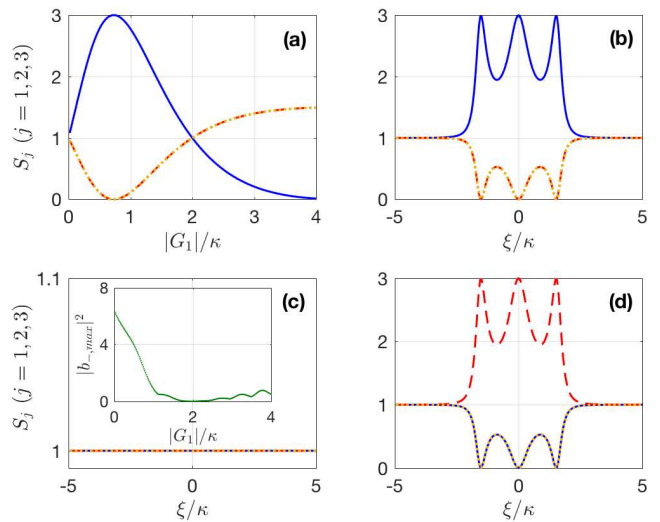} \caption{(Color online) Normalized output energies $S_{1}$ (blue solid), $S_{2}$ (red dashed), and $S_{3}$ (yellow dotted) versus $|G_{1}|$ with $\xi=0$ (a), versus $\xi$ with $|G_{1}|=0.75\kappa$ (b), versus $\xi$ with $|G_{1}|=2\kappa$ (c), and versus $\xi$ with $|G_{2}|=0.75\kappa$ (d). The inset in (c) depicts the maximal excitation number $|b_{-,\text{max}}|^{2}$ versus $|G_{1}|$. We assume $\{|G_{2}|,\,\theta_{1},\,\theta_{2}\}=\{\kappa,\,\pi,\,0\}$ in (a)-(c) and $\{|G_{1}|,\,\theta_{1},\,\theta_{2}\}=\{\kappa,\,0,\,\pi\}$ in (d). Other parameters are $\gamma_{m}=10^{-3}\kappa$ and $G=\kappa$.}
\label{fig5}%
\end{figure}

Instead of focusing on the cumbersome analytical expressions in Eq.~(\ref{eq13}), we plot in Fig.~\ref{fig5} the numerical results to show coherent photon routing. To achieve this in a more simple way, we first assume $G_{2}=G_{3}$ ($|G_{2}|=G$ and $\theta_{2}=0$) and $\theta_{1}=\pi$. Figure~\ref{fig5}(a) shows the normalized output energies corresponding to resonant signals $(\xi=0)$ in this case, from which we find that $S_{2}(0)$ and $S_{3}(0)$ can be completely suppressed while $S_{1}(0)$ is amplified around $|G_{1}|=0.75\kappa$. As shown in Fig.~\ref{fig5}(b), the three signals undergo CPS and only output from cavity $a_{1}$. Moreover, we can find from Fig.~\ref{fig5}(a) that the three output fields become equal when $|G_{1}|=2\kappa$. Interestingly, all three output fields are frequency-independent with $S_{j}\equiv1$ in this case, as shown in Fig.~\ref{fig5}(c). Such a phenomenon is reminiscent of the frequency-independent perfect reflection (FIPR) proposed in~\cite{DLepl,DLoe}, which is attributed to the destructive interference between different phonon excitation paths (three paths for our system, i.e., $a_{1-,2-,3-}\rightarrow b_{-}$). This can be verified by the inset in Fig.~\ref{fig5}(c), where the maximal excitation number $|b_{-,\text{max}}|^{2}=\text{max}[|b_{-}(\xi)|^{2}]$ of the mechanical anti-Stokes component vanishes around $|G_{1}|=2\kappa$. Figure~\ref{fig5}(d) shows that the synthetic signal can also output only from cavity $a_{2}$ if $G_{1}=G_{3}=G$ and $G_{2}=-0.75G$. In this way, the three-port optomechanical system can serve as a controllable photon router with which the synthetic signal can output from the desired target port.

\begin{figure}[ptb]
\includegraphics[width=8.5cm]{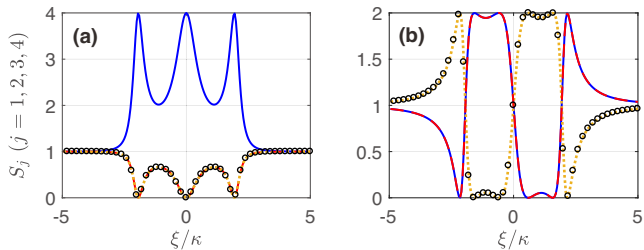} \caption{(Color online) Normalized output energies $S_{1}$ (blue solid), $S_{2}$ (red dashed), $S_{3}$ (yellow dotted), and $S_{4}$ (black circle) versus $\xi$ with (a) $\theta_{1}=\pi$ and $\theta_{2}=0$; (b) $\theta_{1}=\theta_{2}=\pi/2$. Other parameters are $\gamma_{m}=10^{-3}\kappa$, $G=\kappa$, and $\theta_{3}=\theta_{4}=0$.}
\label{fig6}%
\end{figure}

Finally, we reveal that the above results can be extended to a more-port optomechanical system with one control port and more than two target ports. Considering for example a four-port model with an additional cavity mode $a_{4}$ (total decay rate $\kappa_{4}=\kappa_{\text{ex},4}+\kappa_{0,4}$ with $\kappa_{\text{ex},4}$ and $\kappa_{0,4}$ being the external and intrinsic losses, respectively), in which $a_{4}$ is also driven on the red mechanical sideband with the effective optomechanical coupling strength $G_{4}$. We perturb all cavity modes by weak signals of identical amplitude $\varepsilon_{s}$ and vanishing phase. For this system, the anti-Stoles components of the quantum fluctuations can be described by
\begin{subequations}\label{eq16}
\begin{align}
fa_{j-}=&-iGe^{i\theta_{j}}b_{-}+\sqrt{\kappa}\varepsilon_{s},\\
hb_{-}=&-i\sum_{j=1}^{4}Ge^{-i\theta_{j}}a_{j-},
\end{align}
\end{subequations}
where $f=\kappa/2-i\xi$. For simplicity, we have assumed here $\kappa_{\text{ex},j}=\kappa_{j}=\kappa$ and $G_{j}=Ge^{i\theta_{j}}$.

Figure~\ref{fig6} shows the numerical results of the normalized output energies in this case. On one hand, we reveal that the synthetic signal can be totally routed to a desired port $a_{j}$ with $\theta_{j}=\pi$ and $\theta_{j'\neq j}=0$. As shown in Fig.~\ref{fig6}(a) for instance, the signals undergo CPS and output only from cavity $a_{1}$ when $\theta_{1}=\pi$ and $\theta_{2}=\theta_{3}=\theta_{4}=0$. On the other hand, Fig.~\ref{fig6}(b) shows that one can also split the synthetic signal equally into two desired ports: assuming $\theta_{1}=\theta_{2}=\pi/2$ and $\theta_{3}=\theta_{4}=0$, the output spectra become asymmetric with $S_{1(3)}=S_{2(4)}=2$ and $S_{3(1)}=S_{4(2)}=0$ within certain frequency ranges. Such a frequency-dependent light split, with which signals of different frequencies may be splitted to different ports, provides a more flexible scheme of photon routing.

\section{Conclusions}\label{sec5}

In summary, we have proposed a three-port optomechanical system including three indirectly coupled cavity modes and one mechanical mode. While one cavity serves as a control port and is perturbed continually by a control signal, the other two cavities serve as target ports of our interest. Based on the system, two scenarios have been considered under which the system is perturbed in different ways. For the first scenario, we have revealed that the transmission behaviors may be different if another signal is injected upon the two target ports respectively. Such a nonreciprocal phenomenon is attributed to different quantum interferences in opposite directions. In particular, the transmission can be completely suppressed over the full frequency domain, which is referred to as frequency-independent perfect blockade. For the second scenario, all cavity modes are perturbed simultaneously. We have achieved coherent photon routing, with which the synthetic signal only outputs from the desired port. The results can be extended to more-port optomechanical systems, which may provide a feasible scheme of multi-port quantum node and quantum network.

\section*{Acknowledgments}

This work was supported by the Science Challenge Project (Grant No. TZ2018003), the National Key R\&D Program of China (Grant No. 2016YFA0301200), and the National Natural Science Foundation of China (under Grants No. 10534002, No. 11674049, No. 11774024, No. 11534002, No. U1930402, and No. U1730449).

\end{document}